# Self-Dual Vector Multiplet in 3D with Gauged Scale Covariance

Hitoshi Nishino[1] and Subhash Rajpoot[2]

*Department of Physics & Astronomy*
*California State University*
*1250 Bellflower Boulevard*
*Long Beach, CA 90840*

**Abstract**

We present non-trivial interactions of $N=1$ self-dual massive vector multiplet in three-dimensions, with gauged scale covariance. Our multiplets are a vector multiplet $(A_\mu, \lambda)$ and a gauge multiplet $(B_\mu, \chi)$, where the latter is used for the gauging of the scale covariance of the former. Due to the absence of supergravity, this system has no lagrangian formulation, but has only a set of field equations. The gauge multiplet can also have Dirac-Born-Infeld type interactions, even in the presence of the massive self-dual vector multiplet. As a by-product, we also show that scale covariant couplings are possible for scalar multiplet. We also try a mechanism of spontaneous breaking of scale covariance by introducing a superpotential for scalar multiplets.

---

[1]E-Mail: hnishino@csulb.edu

[2]E-Mail: rajpoot@csulb.edu

## 1.  Introduction

The concept of Hodge self-duality was first formulated in even space-time dimensions. This is because if a field strength $F$ of rank $r$ is self-dual in $D$ space-time dimensions, the relationship $F = \star F$ should hold, where the Hodge dual $\star F$ has the rank $D - r$. Then it follows that $r = D - r \implies D = 2r$, namely, $D$ should be an even integer.

However, we can generalize this concept of Hodge self-duality to odd dimensions, as was first shown in [1]. The generalized self-duality is dictated by the relationship $mA = \star F$, where $F$ is the field strength $F = dA$, where $A$ is of rank $r - 1$, while $m$ is a constant with the dimension of mass. Such a self-duality implies the condition $r - 1 = D - r \implies D = 2r - 1$, namely, $D$ is odd. Due to the obvious condition $r - 1 > 0$, the lowest odd dimension for such a self-duality is three-dimensions (3D). In 3D, a vector $A$ was shown to be a massive vector with only one on-shell degree of freedom, even though this vector is massive [1]. This is because the original two on-shell degrees of freedom of massive vector $A$ are halved by the self-duality condition $mA = \star F$. In the original attempts in [1], such generalized self-duality was further supersymmetrized by the addition of a gaugino field $\lambda$. The drawback of such a formulation is the difficulty to couple it to other multiplets, such as supergravity.

Independent of these considerations, any development in supersymmetric theories in 3D has an additional advantage. This is because supersymmetry in 3D has potential application associated with M-theory [2], *via* supermembranes [3]. Therefore any new development in 3D may have a potential impact on M-theory [2] *via* supermembranes [3], if not an immediate application for the time being.

In this paper, we will show that such non-trivial interactions are indeed possible. This will be accomplished by the introduction of scale covariance of a massive vector multiplet (VM). Our scale covariance is different from the conventional dilatation [4], because it commutes with translations. It also differs from the dilatation in conformal supergravity [5], because our scaling weight is common to all the component fields within a given multiplet.

We can couple the $N = 1$ self-dual massive VM to an additional gauge multiplet (GM) $(B_\mu, \chi)$ gauging the scale covariance of the former. We will formulate such a system both in component and superspace languages. As a by-product, we will also show that a similar gauging of scale invariance is possible for a scalar multiplet in 3D. We will also try to break scale covariance spontaneously by a superpotential.

## 2. Component Formulation

We first give a component formulation of our system in the most standard notation. We have two multiplets: The VM $(A_\mu, \lambda_\alpha)$ and the GM $(B_\mu, \chi_\alpha)$. These multiplets are subject to the supersymmetry transformation rule

$$\delta_Q A_\mu = +(\overline{\epsilon}\gamma_\mu \lambda) - m^{-1}(\epsilon \mathcal{D}_\mu \lambda) \ , \tag{2.1a}$$

$$\delta_Q \lambda = +m(\gamma^\mu \epsilon) A_\mu - \tfrac{1}{2}\epsilon(\overline{\chi}\lambda) \ , \tag{2.1b}$$

$$\delta_Q B_\mu = +(\overline{\epsilon}\gamma_\mu \chi) \ , \tag{2.1c}$$

$$\delta_Q \chi = +\tfrac{1}{2}(\gamma^{\mu\nu}\epsilon)G_{\mu\nu} \ , \tag{2.1d}$$

where our 3D metric is $\eta_{\mu\nu} = \text{diag.}(-,+,+)$ with the space-time indices $\mu, \nu, \cdots = 0, 1, 2$ or spinorial indices $\alpha, \beta, \cdots = 1, 2$, while the Clifford algebra is $\{\gamma_\mu, \gamma_\nu\} = +2\eta_{\mu\nu}$. Relevantly, we have $\epsilon^{012} = +1$, $\gamma^{\mu\nu\rho} = \epsilon^{\mu\nu\rho}$, $\gamma^{\mu\nu} = \epsilon^{\mu\nu\rho}\gamma_\rho$, $\gamma^\mu = -(1/2)\epsilon^{\mu\nu\rho}\gamma_{\nu\rho}$. The above rule for the VM is the same as in [1] at the linear level. Our fermionic bilinears are defined by, *e.g.*, $(\overline{\epsilon}\gamma_\mu \lambda) \equiv \epsilon^\alpha (\gamma_\mu)_\alpha{}^\beta \lambda_\beta$. The field strength $G$ is $G_{\mu\nu} \equiv \partial_\mu B_\nu - \partial_\nu B_\mu$. The covariant derivative $\mathcal{D}_\mu$ is scale-covariant under the Abelian scale transformation

$$\delta_\zeta A_\mu = +m\zeta A_\mu \ , \quad \delta_\zeta \lambda = +m\zeta \lambda \ ,$$
$$\delta_\zeta B_\mu = +\partial_\mu \zeta \ , \quad \delta_\zeta G_{\mu\nu} = 0 \ , \quad \delta_\zeta \chi = 0 \ , \tag{2.2}$$

defined by

$$\mathcal{D}_\mu A_\nu \equiv \partial_\mu A_\nu - m B_\mu A_\nu \ ,$$
$$\mathcal{D}_\mu \lambda \equiv \partial_\mu \lambda - m B_\mu \lambda \ . \tag{2.3}$$

Eq. (2.2) is equivalent to the statement that the VM has unit scaling weight, while the GM has zero scaling weight.

We repeat that our scale covariance is different from so-called dilatation [4][5]. This is because our scale transformation commutes with translation, with scaling weights common within a given supermultiplet, as opposed to conventional supersymmetric theories [5].

The existence of the mass parameter $m$ gives a natural reason to introduce such scale covariance, *e.g.*, (2.2) vanishes upon $m \to 0$. The only caveat is that the $m^{-1}$-term in (2.1a) prevents such a limit. Sometimes it is useful to define the 'field strength' of $A$ by

$$\mathcal{F}_{\mu\nu} \equiv \mathcal{D}_\mu A_\nu - \mathcal{D}_\nu A_\mu \ . \tag{2.4}$$

Note, however, that there is *no* gauge covariance for $A_\mu$, such as $\delta_\alpha A_\mu = +\partial_\mu \alpha$. Instead $\mathcal{F}_{\mu\nu}$ is *covariant* but *not invariant* under (2.2):

$$\delta_\zeta \mathcal{F}_{\mu\nu} = +m\zeta \mathcal{F}_{\mu\nu} \ . \tag{2.5}$$

Accordingly, this satisfies the Bianchi identity

$$\mathcal{D}_{[\mu} \mathcal{F}_{\nu\rho]} \equiv -m G_{[\mu\nu} A_{\rho]} \ . \tag{2.6}$$

As these covariant derivatives already show, there are nontrivial interactions between the two multiplets. Note that our system has no lagrangian, because we have no gravity, and therefore no dreibein that could compensate the scale of the lagrangian density. Instead, this system has only a set of field equations:

$$m A_\mu - \tfrac{1}{2} \epsilon_\mu{}^{\rho\sigma} \mathcal{F}_{\rho\sigma} - \tfrac{1}{2} (\overline{\chi} \gamma_\mu \lambda) \doteq 0 \ , \tag{2.7a}$$

$$m \mathcal{D}_\mu A^\mu + \tfrac{1}{2} m \epsilon^{\rho\sigma\tau} G_{\rho\sigma} A_\tau - \tfrac{1}{2} m (\overline{\chi}\lambda) + \tfrac{1}{8} (\overline{\chi} \gamma^{\mu\nu} \lambda) G_{\mu\nu} + \tfrac{1}{2} (\overline{\lambda} \slashed{\partial} \chi) \doteq 0 \ , \tag{2.7b}$$

$$\slashed{\mathcal{D}} \lambda - m \lambda + \tfrac{1}{4} (\gamma^{\mu\nu} \lambda) G_{\mu\nu} + \tfrac{1}{2} m (\gamma^\mu \chi) A_\mu - \tfrac{1}{8} \lambda (\overline{\chi}\chi) \doteq 0 \ , \tag{2.7c}$$

$$\partial_\nu G^{\mu\nu} \doteq 0 \ , \tag{2.7d}$$

$$\slashed{\partial} \chi \doteq 0 \ , \tag{2.7e}$$

where the symbol $\doteq$ stands for a field equation, and (2.7b) is the necessary condition of (2.7a) as the divergence of the latter. Eq. (2.7a) has three terms, where the first two terms are for the usual massive self-duality in 3D [1], while the last term is for an interaction with the GM. If the GM is switched off, there is only the first term $\partial_\mu A^\mu$ in (2.7b) deleting one degree of freedom out of off-shell two degrees of freedom. When the interactions are switched on, there are three additional correction terms in (2.7b). As (2.7c) shows, the $\lambda$-field is massive with one on-shell degree of freedom, balancing that of the massive self-dual vector $A_\mu$ [1]. As has been mentioned earlier, the limit $m \to 0$ is *not* smooth, due to the $m^{-1}$-term in (2.1a). This is also reflected in the interaction terms between the massive VM and GM with *no* $m$.

The confirmation of the field equations (2.7) is not too difficult. There are two crucial steps in the confirmation. The first one is the closure of supersymmetry on $\lambda$, where the necessity of the $\epsilon \chi \lambda$-term in $\delta_Q \lambda$ is revealed. The addition of this term is also related to the necessity of the $\lambda G$-term and $\chi A$-terms in the $\lambda$-field equation (2.7c). The two bosonic field equations (2.7a,b) are obtained by applying a spinorial derivative on the $\lambda$-field equation (2.7c). To be more explicit, applying $\delta_Q$ on (2.7c) yields

$$\delta_Q \left[ \slashed{D}\lambda - m\lambda + \tfrac{1}{4}(\gamma^{\mu\nu}\lambda)G_{\mu\nu} + \tfrac{1}{2}m(\gamma^\mu\chi)A_\mu - \tfrac{1}{8}\lambda(\overline{\chi}\chi) \right]$$

$$= -m\gamma^\mu\epsilon \Big[ mA_\mu - \tfrac{1}{2}\epsilon_\mu{}^{\rho\sigma}\mathcal{F}_{\rho\sigma} - \tfrac{1}{2}(\overline{\chi}\gamma_\mu\lambda)$$

$$\quad + \tfrac{1}{4}m^{-1}\overline{\chi}\gamma_\mu \left\{ \slashed{D}\lambda - m\lambda + \tfrac{1}{4}(\gamma^{\mu\nu}\lambda)G_{\mu\nu} + \tfrac{1}{2}m(\gamma^\mu\chi)A_\mu - \tfrac{1}{8}\lambda(\overline{\chi}\chi) \right\} \Big]$$

$$+ \epsilon \Big[ \left\{ m\mathcal{D}_\mu A^\mu + \tfrac{1}{2}m\epsilon^{\rho\sigma\tau}G_{\rho\sigma}A_\tau - \tfrac{1}{2}m(\overline{\chi}\lambda) + \tfrac{1}{8}(\overline{\chi}\gamma^{\mu\nu}\lambda)G_{\mu\nu} + \tfrac{1}{2}(\overline{\lambda}\slashed{\partial}\chi) \right\}$$

$$\quad - \tfrac{1}{4}\overline{\chi} \left\{ \slashed{D}\lambda - m\lambda + \tfrac{1}{4}(\gamma^{\mu\nu}\lambda)G_{\mu\nu} + \tfrac{1}{2}m(\gamma^\mu\chi)A_\mu - \tfrac{1}{8}\lambda(\overline{\chi}\chi) \right\} \Big] \,, \quad (2.8)$$

which shows nothing but the mutual consistency among (2.7a,b,c,e). Eq. (2.7b) is also confirmed by applying $\mathcal{D}_\mu$ to (2.7a):

$$\mathcal{D}_\mu \left[ mA^\mu - \tfrac{1}{2}\epsilon^{\mu\rho\sigma}\mathcal{F}_{\rho\sigma} - \tfrac{1}{2}(\overline{\chi}\gamma^\mu\lambda) \right]$$

$$= \left[ m\mathcal{D}_\mu A^\mu + \tfrac{1}{2}m\epsilon^{\rho\sigma\tau}G_{\rho\sigma}A_\tau - \tfrac{1}{2}m(\overline{\chi}\lambda) + \tfrac{1}{8}(\overline{\chi}\gamma^{\mu\nu}\lambda)G_{\mu\nu} + \tfrac{1}{2}(\overline{\lambda}\slashed{\partial}\chi) \right]$$

$$\quad - \tfrac{1}{4}\overline{\chi} \left[ \slashed{D}\lambda - m\lambda + \tfrac{1}{4}(\gamma^{\mu\nu}\lambda)G_{\mu\nu} + \tfrac{1}{2}m(\gamma^\mu\chi)A_\mu - \tfrac{1}{8}\lambda(\overline{\chi}\chi) \right] \,, \quad (2.9)$$

consistently with (2.7c,e).

In (2.7d,e) we have given free field equations for the GM as our *simplest choice*, because the consistency among field equations do *not* 'specify' any interactions on the r.h.s. of these equations. Such free field equations look rather strange at first glance, because the VM undergoes interactions with the GM. However, if there were interaction terms with any fields in the VM, there would arise a contradiction, because such terms would be no longer scale invariant, while the l.h.s. is invariant, because $G_{\mu\nu}$ and $\chi$ are scale invariant as in (2.2).

However, the GM can have its own self-interactions. In other words, there can be non-trivial self-interactions on the r.h.s. of (2.7d,e) without upsetting consistency with (2.7a,b,c). A good example is Dirac-Born-Infeld (DBI) type interactions [6][7]. In such a case, eqs. (2.7a,b,c) are intact, while (2.7d,e) are respectively modified as

$$\partial_\nu G^{\nu\mu} + \tfrac{1}{2}\alpha^2 G^{\mu\nu}G^{\rho\sigma}\partial_\nu G_{\rho\sigma} - \tfrac{1}{2}\alpha^2(\partial_\mu\overline{\chi})(\partial_\nu\chi)\widetilde{G}^\nu$$

$$\quad - \tfrac{1}{2}\alpha^2(\overline{\chi}\gamma_\rho\partial_\sigma\chi)\partial^\sigma G^{\mu\rho} + \tfrac{1}{2}\alpha^2(\overline{\chi}\partial_\mu\partial_\nu\chi)\widetilde{G}^\nu + \mathcal{O}(\alpha^4) \doteq 0 \,, \quad (2.10a)$$

$$\slashed{\partial}\chi + \tfrac{1}{4}\alpha^2\chi(\partial_\mu\overline{\chi})(\partial^\mu\chi) - \tfrac{1}{2}\alpha^2(\gamma_\rho\partial_\sigma\chi)G_{\rho\tau}G_\sigma{}^\tau - \tfrac{1}{4}\alpha^2(\gamma^\mu\chi)G_{\rho\sigma}\partial_\mu G^{\rho\sigma} + \mathcal{O}(\alpha^4) \doteq 0 \,, \quad (2.10b)$$

where $\widetilde{G}^\mu \equiv (1/2)\epsilon^{\mu\rho\sigma}G_{\rho\sigma}$. We also use the expression, *e.g.*, $(\partial_\mu\overline{\chi})(\partial^\mu\chi) \equiv (\partial_\mu\chi^\alpha)(\partial^\mu\chi_\alpha)$, where the parentheses are used to restrict the operation of derivatives. The constant $\alpha$ has the dimension of $m^{-1}$. The purely bosonic terms in (2.10a) agree with the $\mathcal{O}(\alpha^2)$-terms derived from the DBI-terms [6][7]

$$\mathcal{L}_{\text{DBI}} = -\alpha^{-2}\left[\sqrt{\det(\eta_{\mu\nu} + \alpha G_{\mu\nu})} - 1\right] = -\tfrac{1}{4}G_{\mu\nu}^2 + \tfrac{1}{32}\alpha^2(G_{\mu\nu}^2)^2 + \mathcal{O}(\alpha^4) \ . \quad (2.11)$$

The field equations in (2.10) transform to each other up to $\mathcal{O}(\alpha^4)$ under supersymmetry (2.1) which is *not* modified by $\alpha$-corrections. In the confirmation of (2.10a) transforming into (2.10b) under supersymmetry, we can ignore any term at $\mathcal{O}(\alpha^2)$ that can vanish upon the use of the free field equations at $\mathcal{O}(\alpha^0)$. For example, a term $\alpha^2 \epsilon (\partial_\rho \overline{\chi}) \gamma^\tau (\partial\!\!\!/ \chi) G_\tau{}^\rho$ is formally at $\mathcal{O}(\alpha^2)$ but can be dropped, because this term is regarded as $\mathcal{O}(\alpha^4)$ upon the $\chi$-field equation $\partial\!\!\!/ \chi \doteq \mathcal{O}(\alpha^2)$ in (2.10b).

## 3. Reformulation in Superspace

We can reformulate our system in superspace that sometimes has advantages. In contrast to a usual massless VM, we introduce *no* superfield strength for our VM $(A_a, \lambda_\alpha)$.[3] Instead, we introduce just the potential superfield $A_A$ and a spinor superfield $\lambda_\alpha$. As for the GM $(B_a, \chi_\alpha)$, we have the superfield strength $G_{AB}$ as usual. The supercovariant and scale covariant derivative $\nabla_A$ is defined by

$$\nabla_A \equiv D_A - m B_A \mathcal{S} \ , \quad (3.1a)$$

$$[\nabla_A, \nabla_B\} = T_{AB}{}^C \nabla_C - m G_{AB} \mathcal{S} \ , \quad (3.1b)$$

where the supercovariant derivative $D_A$ is defined by

$$D_a = \partial_a \ , \quad D_\alpha \equiv \partial_\alpha - (\gamma^b \theta)_\alpha \partial_b \ . \quad (3.2)$$

Here $D_A$ is the usual supercovariant derivative before gauging [8], while $\mathcal{S}$ is the generator for scale covariance, acting as

$$\mathcal{S} A_A = +A_A \ , \quad \mathcal{S} \lambda_\alpha = +\lambda_\alpha \ , \quad \mathcal{S} G_{AB} = 0 \ , \quad \mathcal{S} \chi_\alpha = 0 \ . \quad (3.3)$$

The $\nabla_a$ corresponds to the component covariant derivative $\mathcal{D}_\mu$. Accordingly, the $G$-Bianchi identities are

$$\nabla_{[A} G_{BC)} - T_{[AB|}{}^D G_{D|C)} \equiv 0 \ . \quad (3.4)$$

---

[3]Our notation in superspace is different from component ones. We use $a, b, \cdots = 0, 1, 2$ for the space-time vector indices, while $\alpha, \beta, \cdots = 1, 2$ for spinorial indices. The antisymmetrization in superspace is such as $S_{[A} T_{B)} \equiv S_A T_B \pm S_B T_A$ with *no* factor $1/2$ in front.

Our superspace constraints consistent with our previous component results are

$$T_{\alpha\beta}{}^c = +2(\gamma^c)_{\alpha\beta} ~, \tag{3.5a}$$

$$T_{\alpha b}{}^D = 0 ~, \quad T_{\alpha\beta}{}^\delta = 0 ~, \quad T_{ab}{}^D = 0 ~, \tag{3.5b}$$

$$G_{\alpha b} = +(\gamma_b)_\alpha{}^\beta \lambda_\beta \equiv +(\gamma_b \lambda)_\alpha ~, \quad G_{\alpha\beta} = 0 ~, \tag{3.5c}$$

$$\nabla_\alpha \lambda_\beta = -m(\gamma^c)_{\alpha\beta} A_c - \tfrac{1}{2} C_{\alpha\beta}(\overline{\chi}\lambda) ~, \tag{3.5d}$$

$$\nabla_\alpha \chi_\beta = -\tfrac{1}{2}(\gamma^{cd})_{\alpha\beta} G_{cd} ~, \tag{3.5e}$$

where $C_{\alpha\beta}$ is the antisymmetric charge conjugation matrix in 3D. Our field equation (2.7) in component can be recast into superspace language as

$$mA_a - \tfrac{1}{2}\epsilon_a{}^{bc}\mathcal{F}_{bc} - \tfrac{1}{2}(\overline{\chi}\gamma_a\lambda) \doteq 0 ~, \tag{3.6a}$$

$$m\nabla_a A^a + \tfrac{1}{2}m\epsilon^{abc} G_{ab} A_c - \tfrac{1}{2}m(\overline{\chi}\lambda) + \tfrac{1}{8}(\overline{\chi}\gamma^{ab}\lambda)G_{ab} + \tfrac{1}{2}(\overline{\lambda}\nabla\!\!\!/\chi) \doteq 0 ~, \tag{3.6b}$$

$$\left[ +\nabla\!\!\!/\lambda - m\lambda + \tfrac{1}{4}\gamma^{ab}\lambda G_{ab} + \tfrac{1}{2}m(\gamma^a\chi)A_a - \tfrac{1}{8}\lambda(\overline{\chi}\lambda) \right]_\alpha \doteq 0 ~, \tag{3.6c}$$

$$\nabla_b G^{ab} \doteq 0 ~, \tag{3.6d}$$

$$(\nabla\!\!\!/\chi)_\alpha \doteq 0 ~. \tag{3.6e}$$

Even though we used the notation $\mathcal{F}$, it stands just for $\mathcal{F}_{ab} \equiv \nabla_a A_b - \nabla_b A_a$, but it has *no* corresponding Bianchi identity. This is because there is *no* actual gauge symmetry for the massive gauge field $A$ starting with $\delta_\alpha A_a = \nabla_a \alpha$.

The confirmation of these constraints and field equations goes as follows. As for the Bianchi identities (3.4), the procedure is the usual routine, starting with the dimension $d = 1/2$ going up to $d = 2$. However, since the potential superfield $A_A$ has no gauge invariance with *no* Bianchi identities, we have to confirm the commutator $[\nabla_A, \nabla_B\}A_c$ up to $d = 2$.

Corresponding to (2.10), we can introduce the DBI-type interactions [7] into the GM, replacing (3.6d,e). Eq. (2.10) is recast into superspace form replacing (3.6d,e):

$$\nabla_b G^{ba} + \tfrac{1}{2}\alpha^2 G^{ab} G^{cd} \nabla_b G_{cd} - \tfrac{1}{2}\alpha^2 (\nabla_a\overline{\chi})(\nabla_b\chi)\widetilde{G}^b$$
$$- \tfrac{1}{2}\alpha^2(\overline{\chi}\gamma^c\nabla_d\chi)\nabla^d G^{ac} + \tfrac{1}{2}\alpha^2(\overline{\chi}\nabla_a\nabla_b\chi)\widetilde{G}^b + \mathcal{O}(\alpha^4) \doteq 0 ~, \tag{3.7a}$$

$$\nabla\!\!\!/\chi + \tfrac{1}{4}\alpha^2\chi(\nabla_a\overline{\chi})(\nabla^a\chi) - \tfrac{1}{2}\alpha^2(\gamma_c\nabla_d\chi)G^{cd}G_d{}^e - \tfrac{1}{4}\alpha^2(\gamma^a\chi)G_{cd}\nabla_a G^{cd} + \mathcal{O}(\alpha^4) \doteq 0 ~. \tag{3.7b}$$

## 4. Application to a Scalar Multiplet

The success of gauging scale covariance for a self-dual massive VM motivates us to apply the results to a scalar multiplet (SM) in 3D. This is based on the viewpoint that the massive VM is essentially equivalent to a SM with $1+1$ on-shell degrees of freedom.

In this section, we introduce two multiplets: The SM $(A, \psi_\alpha, F)$ and a GM $(B_a, \chi_\alpha)$. The former has $1+1$ on-shell degrees of freedom, while $2+2$ off-shell with the auxiliary field $F$. We formulate the system in terms of $N=1$ superspace for simplicity. For the same reason for the previous self-dual massive VM, the system has no lagrangian formulation, but a set of field equations.

As before, scale covariance will be gauged, so that the basic supercovariant derivative and $G$-Bianchi identities are the same as in (3.1) and (3.4). We introduce a real scalar superfield $\Phi$ whose $\theta$ sectors define the component fields as

$$\Phi| = A \ , \quad \nabla_\alpha \Phi| = \psi_\alpha \ , \quad \nabla^2 \Phi| = F \ , \tag{4.1}$$

where the vertical bars stand for restrictions to the $\theta = 0$ sectors, and $\nabla^2 \equiv (1/2)\nabla^\alpha \nabla_\alpha$ as in [8]. We assign the unit scaling weight to our SM: $\mathcal{S}\Phi = +\Phi$. As for the GM, the $G$-Bianchi identities are the same as (3.1) with the constraints (3.5a,b,c). The SM field equations are controlled only by a single superfield equation

$$\nabla^2 \Phi - \mu \Phi \doteq 0 \ , \tag{4.2}$$

where $\mu$ is a mass parameter generally distinct from $m$. Eq. (4.2) is equivalent to the set of component field equations

$$\nabla_a^2 A - \mu F + m\lambda^\alpha \psi_\alpha \doteq 0 \ , \tag{4.3a}$$

$$(\nabla\!\!\!\!/\,\psi)_\alpha - \mu\psi_\alpha - m\lambda_\alpha A \doteq 0 \ , \tag{4.3b}$$

$$F - \mu A \doteq 0 \ . \tag{4.3c}$$

The mutual consistency of equations in (4.3) can be confirmed by applying spinorial derivatives on each field equations, as usual [8].

Interestingly, in this process we realize that the GM field equations are *not* necessarily required to be the free field equations (3.6d,e). In other words, we can here again introduce non-trivial DBI-type self-interactions of the GM [6][7]. In such a case, the field equations of $\chi$ and $B$ will be exactly the same as (3.7). Needless to say, we can have both a VM and a SM coupled to a GM at the same time.

## 5. Trial of Spontaneous Breaking of Scale Covariance

In this section, we try a spontaneous breaking of scale covariance which is the next natural step to consider. For that purpose we need some superpotential to give nontrivial v.e.v.s to certain scalar fields. However, it is easily seen that just a single SM can not do the job, because any polynomial function of a SM automatically breaks the original scale covariance at the free-field level.

To solve this problem, we introduce a pair of SMs represented by the real scalar superfields $\Phi$ and $\widetilde{\Phi}$. The trick here is to assign opposite scaling weights for these superfields:

$$\mathcal{S}\Phi = +\Phi \ , \qquad \mathcal{S}\widetilde{\Phi} = -\widetilde{\Phi} \ , \tag{5.1}$$

so that the product $\Phi\widetilde{\Phi}$ carries zero scaling weight, allowing to form any polynomial functions for a superpotential.

Based on this preliminary, our superfield equations will be

$$\nabla^2 \Phi + c\Phi f'(\Phi\widetilde{\Phi}) \doteq 0 \ , \tag{5.2a}$$

$$\nabla^2 \widetilde{\Phi} + c\widetilde{\Phi} f'(\Phi\widetilde{\Phi}) \doteq 0 \ , \tag{5.2b}$$

$$D^\beta D_\alpha W_\beta + \widehat{c}\, m(\Phi \nabla_\alpha \widetilde{\Phi} - \widetilde{\Phi} \nabla_\alpha \Phi) \doteq 0 \ , \tag{5.2c}$$

where $c$ and $\widehat{c}$ are arbitrary nonzero real constants. The current conservation for scale transformation, *i.e.*, the $D^\alpha$-operation on (5.2c) requires that the last terms in (5.2a,b) should have the common constant $c$. The real function $f(\xi)$ of a real number $\xi$ is *a priori* arbitrary real function. However, for the system to be renormalizable, we choose it to be at most bilinear in $\xi$:

$$f(\xi) \equiv f_0 - \mu\xi + \tfrac{1}{2}g\xi^2 \ , \qquad f'(\xi) = -\mu + g\xi \ , \qquad f''(\xi) = +g \ , \tag{5.3}$$

where $f_0$, $\mu$ and $g$ are arbitrary constants. All other algebras related to the GM are the same as in section three.

We now analyze the component field equations of this system:

$$F + cA(-\mu + gA\widetilde{A}) \doteq 0 \ , \qquad \widetilde{F} + c\widetilde{A}(-\mu + gA\widetilde{A}) \doteq 0 \ , \tag{5.4a}$$

$$\not{\nabla}\chi - m\lambda A + c\chi f'(A\widetilde{A}) + cgA(\chi\widetilde{A} + \widetilde{\chi}A) \doteq 0 \ , \tag{5.4b}$$

$$\not{\nabla}\widetilde{\chi} - m\lambda\widetilde{A} + c\widetilde{\chi} f'(A\widetilde{A}) + cg\widetilde{A}(\chi\widetilde{A} + \widetilde{\chi}A) \doteq 0 \ , \tag{5.4c}$$

$$\nabla_a^2 A + m(\overline{\lambda}\chi) + c\left[ Ff'(A\widetilde{A}) + gA\widetilde{A}F + gA^2\widetilde{F} \right] + 2cg\left[ \widetilde{A}\chi^2 + \widetilde{A}(\overline{\chi}\widetilde{\chi}) \right] \doteq 0 \ , \tag{5.4d}$$

$$\nabla_a^2 \widetilde{A} + m(\overline{\lambda}\widetilde{\chi}) + c\left[\widetilde{F} f'(A\widetilde{A}) + gA\widetilde{A}\widetilde{F} + g\widetilde{A}^2 F\right] + 2cg\left[A\widetilde{\chi}^2 + \widetilde{A}(\overline{\chi}\chi)\right] \doteq 0 \ , \quad (5.4\text{e})$$

$$\nabla_b F^b{}_a - \tfrac{1}{2}\widehat{c}\, m(A\nabla_a\widetilde{A} - \widetilde{A}\nabla_a A) - \tfrac{1}{2}\widehat{c}\, m(\overline{\chi}\gamma_a\widetilde{\chi}) \doteq 0 \ , \quad (5.4\text{f})$$

$$\not{\nabla}\lambda + \tfrac{1}{2}\widehat{c}\, m(A\widetilde{\chi} - \widetilde{A}\chi) \doteq 0 \ . \quad (5.4\text{g})$$

We have omitted the spinorial indices for fermionic field equations (5.4b,c,g).

We next study the possible spontaneous symmetry breaking. To this end, we first ignore all the fermionic fields and space-time derivatives in all the field equations. We next eliminate the $F$ and $\widetilde{F}$-fields, getting the algebraic field equations

$$c^2 A(\mu - gA\widetilde{A})(\mu - 3gA\widetilde{A}) \doteq 0 \ . \quad (5.5)$$

As usual in supersymmetric models, the supersymmetric vacuum configuration with $\langle F\rangle = \langle\widetilde{F}\rangle = 0$ corresponds to the solutions

$$\langle A\widetilde{A}\rangle = g^{-1}\mu \ , \quad \langle F\rangle = 0 \ , \quad \langle\widetilde{F}\rangle = 0 \ . \quad (5.6)$$

Relevantly, we see that the $B$-field equation is

$$\nabla_b F^b{}_a - \widehat{c}\, m^2 \langle A\widetilde{A}\rangle A_a + \mathcal{O}(\phi^2) \doteq 0 \ , \quad (5.7)$$

up to some interaction terms $\mathcal{O}(\phi^2)$. This means that the $B$-field acquires the mass $\sqrt{\widehat{c}}\, m$, as desired for a breaking of scale covariance.

The only drawback of this model is that the vacuum configuration (5.6) is not stable. In fact, the bosonic interaction terms of the $A$ and $\widetilde{A}$-field equations

$$\nabla_a^2 A - c^2 A(\mu - gA\widetilde{A})(\mu - 3gA\widetilde{A}) \doteq 0 \ , \quad (5.8\text{a})$$

$$\nabla_a^2 \widetilde{A} - c^2 \widetilde{A}(\mu - gA\widetilde{A})(\mu - 3gA\widetilde{A}) \doteq 0 \ , \quad (5.8\text{b})$$

are integrable to yield the bosonic potential

$$V = +c^2 A\widetilde{A}(\mu - gA\widetilde{A})^2 \ . \quad (5.9)$$

This form seems 'almost' positive definite, but it is not actually because of the factor $A\widetilde{A}$ which can be negative. In other words, this system has a potential unbounded from below. This is also traced back to the combination of $A\widetilde{A}$.

We also mention the possibility with non-renormalizable interactions. In such a case, the potential in (5.9) might be positive definite, since more general factors, such as $(A\widetilde{A})^2$ instead of $A\widetilde{A}$ can be allowed.

In any case, despite the above unstable configuration, our result suggests other possibilities yet to be explored in the future, because breaking of scale covariance may become also important for applications associated with supermembrane physics in 3D [3].

## 6. Concluding Remarks

In this paper, we have shown that the massive self-dual VM [1] with $N=1$ supersymmetry in 3D can be coupled to a GM which gauges the scale covariance of the former. The existence of the mass parameter provides a natural reason to have a scale covariance of the massive multiplet. Our scale transformation differs from the conventional dilatation [4] in conformal supersymmetry [5]. As a by-product, we have also shown that similar scale covariant couplings are possible for a SM in 3D. We also tried a spontaneous breaking of scale covariance by introducing a superpotential.

The result of this paper provides two important significances. First, the consistent couplings of massive self-dual VM [1] with a GM are shown to be possible. Second, there can be a scale covariance for the massive self-dual VM, and moreover it can be gauged by an additional GM consistently in superspace. To our knowledge, there has been no such a result in the past, in any dimensions with supersymmetry, not to be limited to 'self-dual' massive VM.

It is interesting that even a singlet VM with no gauge index can acquire a scale covariance which can be gauged consistently in superspace. Since a similar massive VM exists also in 9D with 8+8 degrees of freedom with $N=1$ supersymmetry [9], it seems interesting to consider such couplings in 9D. The difference is that the massive VM is no longer self-dual as opposed to the 3D case [1] we have treated here. However, we easily see that the GM $(B_a, \chi_\alpha, \varphi)$ in 9D [10], has an extra scalar $\varphi$ that prevents similar couplings. This shows up first in the commutator $\{\nabla_\alpha, \nabla_\beta\} A_c$, because $G_{\alpha\beta} \approx \delta_{\alpha\beta}\varphi$ produces a term $\varphi A_c$ with no counter-term to cancel in the commutator. A similar situation exists in 5D for massive VM with $4+4$ degrees of freedom, with a GM with $4+4$. In this sense, 3D is very special, because both massive VM and GM have no extra scalar that would have prevented nontrivial couplings between these two multiplets.

We have shown in a recent paper [11] that scale invariance may well be playing an important role even in the standard model in 4D. Here in the present paper, we have shown another important role played by scale covariance in 3D. Namely, a vector here can be *covariant* under scale transformations, while it is *invariant* in [11].

The success of gauging scale covariance in 3D encourages us to investigate similar possibilities in higher-dimensions, even independent of the 'massiveness' of VMs. As a matter of fact, we have found that such a formulation is also possible in 10D [12]. We have found not only $N=1$ supersymmetric scale covariance in 10D, but also $N=4$ in 4D by dimensional reduction whose interactions have never been known before.

We are grateful to W. Siegel for important discussions and reading the draft. This work is supported in part by NSF Grant # 0308246.

# References

[1] P.K. Townsend, K. Pilch and P. van Nieuwenhuizen, Phys. Lett. **136B** (1984) 38; Addendum: **137B** (1984) 443.

[2] *See, e.g.,* C. Hull and P.K. Townsend, Nucl. Phys. **B348** (1995) 109; E. Witten, *ibid.* **443** (1995) 85; P.K.Townsend, in *Proceedings of ICTP Summer School on High Energy Physics and Cosmology*, Trieste, 1996, hep-th/9612121; *'M-theory from its Superalgebra'*, hep-th/9712004; *and references therein.*

[3] E. Bergshoeff, E. Sezgin and P.K. Townsend, Phys. Lett. **189B** (1987) 75; Ann. of Phys. **185** (1988) 330.

[4] F. Gürsey, Nuovo Cim. **3** (1956) 988; H.A. Kastrup, Ann. of Phys. **7** (1962) 388; Phys. Rev. **D142** (1966) 1060; *ibid.* **143** (1966) 1041; *ibid.* **150** (1966) 1189; Nucl. Phys. **58** (1964) 561; T. Fulton, R. Rohrich and L. Witten, Rev. Mod. Phys. **34** (1962) 442; G. Mack, Nucl. Phys. **B5** (1968) 499; G. Mack and A. Salam, Ann. of Phys. **53** (1969) 174.

[5] J. Wess and B. Zumino, Nucl. Phys. **B70** (1974) 39; M. Kaku, P.K. Townsend and P. van Nieuwenhuizen, Phys. Rev. **D17** (1978) 3179; J.C. Romao, A. Ferber and P.G.O. Freund, Nucl. Phys. **B122** (1977) 170; S. Ferrara, M. Kaku and P. van Nieuwenhuizen, Nucl. Phys. **B129** (1977) 125; P.S. Howe and R.W. Tucker, Phys. Lett. **80B** (1978) 138; W. Siegel, Phys. Lett. **80B** (1979) 224; S.J. Gates, Jr., Nucl. Phys. **B162** (1980) 79.

[6] M. Born and L. Infeld, Proc. Roy. Soc. Lond. **A143** (1934) 410; *ibid.* **A144** (1934) 425; P.A.M. Dirac, Proc. Roy. Soc. Lond. **A268** (1962) 57.

[7] E. Bergshoeff, M. Rakowski and E. Sezgin Phys. Lett. **185B** (1987) 371.

[8] S.J. Gates, Jr., M.T. Grisaru, M. Roček and W. Siegel, *'Superspace'* (Benjamin/Cummings, Reading, MA 1983).

[9] *See, e.g.,* J. Strathdee, Int. Jour. Mod. Phys. **A2** (1987) 273.

[10] *See, e.g.,* S.J. Gates, Jr., H. Nishino and E. Sezgin, Class. and Quant. Gr. **3** (1986) 21.

[11] H. Nishino and S. Rajpoot, *'Broken Scale Invariance in the Standard Model'*, CSULB-PA-04-02, hep-th/0403039.

[12] H. Nishino and S. Rajpoot, *'Supersymmetric Gauged Scale Covariance in Ten and Lower Dimensions'*, CSULB-PA-04-6, hep-th/0407203.